\documentclass[letterpaper]{JHEP3}
\usepackage{amsmath,amssymb,graphicx,subfigure}

\newcommand{\be}{\begin{equation}}
\newcommand{\ee}{\end{equation}}
\newcommand{\bea}{\begin{eqnarray}}
\newcommand{\eea}{\end{eqnarray}}

\renewcommand{\b}[1]{\overline{#1}}

\newcommand{\ds}{de~Sitter}
\newcommand{\sbb}{\ensuremath{S_{BB}}}
\newcommand{\tbb}{\ensuremath{t_{BB}}}
\newcommand{\bbb}{\ensuremath{B_{BB}}}
\newcommand{\bb}{Boltzmann Brain}
\newcommand{\dtb}{{\ensuremath{\b{D3}}}}
\newcommand{\htip}{{\ensuremath{h_{\rm tip}}}}
\newcommand{\kah}{{\ensuremath{{\cal K}}}}

\title{Evidence for a bound on the lifetime of de Sitter space}

\author{Ben Freivogel\\
Department of Physics and Center for Theoretical
Physics \\
University of California, Berkeley, CA 94720, U.S.A. \\
{\em and}\\
Lawrence Berkeley National Laboratory, Berkeley, CA 94720, U.S.A.\\
\email{freivogel@berkeley.edu}}

\author{Matthew Lippert\\
Department of Physics \\
Technion, Haifa 32000, Israel \\
{\em and} \\
Department of Mathematics and Physics \\
University of Haifa at Oranim, Tivon 36006, Israel\\
\email{matthewslippert@gmail.com}}
 
 \date{}

\abstract
{Recent work has suggested a surprising new upper bound on the lifetime of de Sitter
vacua in string theory.  The bound is parametrically longer
than the Hubble time but parametrically shorter than the
recurrence time. We investigate whether the bound is satisfied in a particular 
class of de Sitter solutions, the KKLT vacua. 
Despite the
freedom to make the supersymmetry breaking scale exponentially small, which naively
would lead to extremely stable vacua,
we find that the lifetime is always less than about $\exp(10^{22})$ Hubble
times, in agreement with the proposed bound.  }

\begin{document}

\section{Introduction}
\label{intro}
String theory appears to contain a large number of \ds\ vacua. Our
current understanding is that \ds\ vacua cannot be completely stable \cite{kklt, dks},
necessarily decaying before the Poincare recurrence time,
\be
t_{\rm rec} \sim H^{-1} e^{S_{dS}}
\ee
where $S_{dS}$ is the entropy of the cosmological horizon,
\be
S_{dS} \sim {M_P^2 \over H^2}
\ee
and $H$ is the Hubble constant. 

Recently, theoretical considerations have suggested a more stringent
bound on the maximum lifetime of \ds\ vacua in string theory \cite{bf}.
As we will explain in section 2, the bound comes from demanding that one causal patch of \ds\
space does not contain an enormous number of observers formed from
rare processes which violate the second law of thermodynamics.  The
bound is much longer than the Hubble time but much shorter than the
recurrence time. The bound is 
\be
t_{\rm decay} < H^{-1} ~ e^{\displaystyle 10^{40}}~.
\ee
As we will explain later, significant theoretical uncertainty remains
in this bound. We estimate that at $1\sigma$ the bound is 
\be
t_{\rm decay} < H^{-1} ~ e^{\displaystyle 10^{40 \pm 20}}~.
\ee
With an uncertainty of 20 in the second exponent, this may be the
least precise prediction in the history of science.
 
Nevertheless, the bound is nontrivial and unexpected 
from the point of view of low energy effective
field theory. Consider gravity coupled to a single scalar field whose
potential contains one minimum with positive vacuum energy and one
minimum with  negative vacuum energy. For a
high, wide barrier, the decay time is of order the recurrence
time. For a low, narrow barrier the decay time is much faster than the
recurrence time. Both situations are robust against corrections, and
from the low energy point of view there seems to be no reason to
consider one class of potentials and not the other \cite{abj, bfl}.

We attempt to construct vacua with such long lifetimes in string
theory, focusing on the construction of Kachru, Kallosh, Linde, and Trivedi (KKLT) \cite{kklt}. 
Since the KKLT scenario allows for
a very low supersymmetry breaking scale, and supersymmetry guarantees
stability, at first it may seem easy to construct extremely stable
vacua. For example, Ceresole, Dall'Agata, Giryavets, Kallosh, and
Linde 
\cite{cdgkl} estimated in a particular context
that the lifetime of nearly supersymmetric vacua is of order 
\be
t_{\rm decay} \sim \exp \left({M_P^2 \over m_{3/2}^2}\right)~.
\ee
(Here and below, we do not compute the one-loop determinant, so the
dimensional prefactor factor in all of our decay times will be unkown.)
While we agree with their analysis in the context it was done and will make use of it later, 
we find that the above formula overestimates the lifetime of
 KKLT vacua with very low supersymmetry breaking scale. 
 
 Instead, we
find that as the supersymmetry breaking scale is lowered the lifetime
approaches a finite limit. We find
\be
t_{\rm decay} < \exp \left( 3 \cdot 10^{-3} {g_s M^6 \over (N_{\dtb})^3} \right)
\ee
where $M$ is a flux number and $N_\dtb$ is the number of anti-$D3$ branes, even though
the supersymmetry breaking scale is exponentially small,
\be
m_{3/2} \sim \exp \left(- { 2 \pi K \over 3 g_s M} \right)
\ee
where $K$ is another flux number.
Tadpole cancellation bounds the flux numbers by the Euler number of the Calabi-Yau
fourfold, so we can bound the lifetime by
\be
t_{\rm decay} < \exp \left( 10^{-9} \chi^5 \right)
\ee
Assuming that the Euler number of Calabi-Yau fourfolds is bounded, and that the
bound is of order the maximum known Euler number, we get a numerical bound
\be
t_{\rm decay} < \exp \left( 10^{22} \right)
\ee
Clearly our result is highly sensitive to the maximum Euler number.

The intuitive explanation for why the lifetime is insensitive to the supersymmetry
breaking scale is the following. Recall that KKLT break
supersymmetry by adding an anti-$D3$ brane at the tip of a warped
throat. The supersymmetry breaking scale can be exponentially low due
 to the exponential redshift in the throat. The decay of the
 nonsupersymmetric \ds\ vacuum is described by an NS5 brane wrapping a
 3-sphere at the tip of the throat. In the 4-dimensional description,
 the wrapped NS5 brane is a domain wall. What happens is that although the
 SUSY breaking scale is exponentially small, the very same warp factor
 guarantees that the tension of the
 domain wall is also exponentially small. We will see that these two
 warp factors cancel in computing the decay rate for long throats. In
 other words, the decay is a process
 localized near the tip of the throat, and so the rate is actually
 insensitive to the length of the throat for sufficiently long throats.
 
 We are focusing on a tiny piece of the string theory landscape. We urge other authors to 
try to construct extremely stable vacua using other constructions, because our results may be highly model dependent. We present here one small piece of evidence that the surprising bound demanded by Boltzmann Brain considerations may actually be obeyed by the landscape of string theory.

Recent work on the lifetimes of string theory vacua includes
interesting papers by Westphal \cite{westphal}, by Dine and
collaborators \cite{dfmb}, and by Johnson and Larfors \cite{jl}. 
These authors, however, were concerned with
stability on time scales of order the Hubble
time. Here we focus in on one corner of the landscape and investigate
a new time scale.

We begin, in section \ref{boltzmannbrain}, with a discussion of Boltzmann Brains to motivate the 
need for a bound on the lifetimes of de Sitter vacua.  In section \ref{CDL_review} we review the physics of false vacuum decay, reminding the reader that at this level it is not difficult to construct false vacua which live for about the recurrence time.  Section \ref{rate_simple} presents a calculation of the decay rate using the brane description of the instanton, while in section \ref{corrections} we consider corrections to the tension of the domain wall due to closed string moduli.  In section \ref{KKLTproblems} we point out the difficulty of constructing de Sitter vacua using the KKLT method. In particular, we show that there is only a narrow window where the construction is marginally under control. However, it is possible that these difficulties can be easily fixed by minor modifications of the KKLT construction. We conclude in section \ref{conclusions}.

\section{The Boltzmann Brain problem}
\label{boltzmannbrain}

String theory appears to contain a vast landscape of stable and
metastable vacua. What we normally think of as constants of nature,
such as the cosmological constant and the electron mass, vary from one
vacuum to another. String theory also appears to contain a mechanism
for producing large regions of spacetime in each one of these vacua:
eternal inflation.

In the eternally inflating multiverse,
intelligent observers form in many different regions. Different
observers will see different cosmological constants, different
electron masses, and different CMB multipoles. 
In this setting, theoretical predictions for the results of experiments are
necessarily statistical \cite{bp, susskind}.
The probability of a given experimental outcome is proportional to the
number of observations, in the multiverse, of that outcome.

Many problems remain in making this framework precise. One is that we
have not precisely defined what constitutes an observation. Another is
that the entire formulation so far relies on the semiclassical
approximation. But, even if we work in the semiclassical approximation
and take some definition of an observation, our ability to make
predictions is hindered by a familiar hobgoblin of theoretical
physics: a problem of infinities.

Eternal inflation produces an infinite volume of spacetime, an
infinite number of ``pocket universes'' of each type, and an infinite
number of observers inside {\it each} pocket universe. Different
seemingly natural prescriptions for regulating the infinities lead to
drastically different predictions. A prescription for regulating infinities and 
extracting predictions is referred to as a {\it measure}.

Fortunately, most simple prescriptions lead to predictions in sharp
conflict with observation. One test of a measure is the 
``Boltzmann Brain problem'' \cite{dks, page, bf}.
There are two basic ways in which structure can form. It can form in the
usual way via inflation, reheating, and gravitational
collapse. Structure can also form through rare thermal fluctuations
which decrease the entropy. For example, a diffuse gas of particles
can spontaneously form a planet populated by intelligent observers. We
will refer to observers produced in the usual way as ``ordinary
observers,'' and observers produced by rare thermal fluctuations as
``Boltzmann Brains.''

Our observations indicate that we are ordinary observers. The reason
is that when structure forms by rare thermal fluctuations, the second
law of thermodynamics is violated. The probability of a rare fluctuation is supressed by
the amount of second law violation, $P \sim \exp(\Delta S)$. So
fluctuating a large, homogeneous universe full of structure is exponentially rarer than
fluctuating a small amount of structure. On the other hand, the number
of observers produced is only proportional to $\Delta S$. Observers
who form from rare thermal fluctuations do not see stars in the
sky. In fact, with a particular definition of what constitutes an
observer, the typical observer formed by thermal fluctuations is an
isolated brain in empty space, which just lives long enough to realize
it exists -- a Boltzmann Brain. We will not need to refer to such
extreme limits here and use the term ``Boltzmann Brain" to refer to any
observer which forms as a result of second law violation.

\subsection{Boltzmann Brains in our causal patch}

To get used to this strange idea, let us first discuss Boltzmann
Brains within our horizon. As far as we know,
our vacuum may have a lifetime of order the recurrence time. (In section \ref{CDL_review} we review the arguments leading to this conclusion.)
Let us assume for the moment that our vacuum lives for approximately
the recurrence time. What are the consequences? 

We restrict attention
to one causally connected region; the volume of this causal patch is
$H^{-3}$. We want to ask the following question: within one causal
patch, how many Earths form from rare thermal fluctuations
(``Boltzmann Earths''), and how
many Earths form in the usual way (``ordinary Earths'')?

In a system at finite temperature $\beta^{-1}$, the time to produce a fluctuation
of energy $E$ is given by
\be
t \approx \beta e^{\beta E}~.
\ee
where the prefactor is typically of order $\beta$ but can depend on details such
as coupling constants.
In our case, this means that the time to form a Boltzmann Earth is
\be
t_{BE} \approx H^{-1}~ e^{ H^{-1}M_E}
\ee
Plugging in the values, we find
\be
t_{BE} \approx (10^{10}~ {\rm years}) e^{10^{92}}
\ee
Continuing to assume that the lifetime of our vacuum is of order the recurrence time,
the number of Boltzmann Earths produced before our vacuum decays is
\be
N_{BE} = {t_{\rm decay} \over t_{BE}} \approx { H^{-1}e^{10^{123}}\over H^{-1}
  e^{10^{92}} }
\ee
Dividing, we find
\be
N_{BE} \approx e^{10^{123}}
\ee
On the other hand, the number of ordinary Earths in our causal patch
is roughly equal to the number of stars inside our horizon,
\be
N_{OE} \approx 10^{22}~.
\ee
Therefore, assuming that our vacuum lives for about the recurrence
time, we find that our causal patch contains far more Boltzmann Earths
than ordinary Earths,
\be
{N_{BE} \over N_{OE} } = e^{10^{123}}
\ee
It is easy to forget how large double-exponential numbers are, so we write
the ratio as a single exponential
\be
{N_{BE} \over N_{OE} } =
e^{100000000000000000000000000000000000000000000000000000000000000000000000000000000000000000000000000000000000000000000000000}
\ee
except that it will not fit on the page. 
The numbers involved are unimaginably large.

If our vacuum lives for about the recurrence time, the number of
Earths produced by ordinary structure formation is completely
negligible compared to the number produced by rare thermal
fluctuations. Yet, as we discussed above,
observation indicates that our Earth was formed in the ordinary
way. Therefore, if our vacuum lives  for about the recurrence time, 
we are extraordinarily
atypical among civilizations in our causal patch. 

Can we conclude that our vacuum must {\it not} live for the
recurrence time? 
The answer is that we really need a measure to answer this
question. Intuitively, one might expect that it does not matter if our
causal patch is dominated by Boltzmann Brains. After all, it takes a
long time for the Boltzmann Brains to form, and in the meantime
more ordinary observers are produced elsewhere in the multiverse. The
infinities must be regulated before we can definitively say that
comparing the number of Boltzmann Brains to ordinary observers in one
causal patch is a meaningful thing to do.

\subsection{Boltzmann Brains in the Landscape}

More generally, string theory contains a large number of
de Sitter vacua. Above we focused on the production of Boltzmann
Earths in our vacuum, but to compare the number of Boltzmann Brains to the number of ordinary observers in the multiverse we need a more general definition of what constitutes an
``observer." It seems most robust to characterize observers
by requiring them to have a certain complexity. Thus in general we
will characterize Boltzmann Brains as ordered systems with at least a  minimum
number of degrees of freedom \sbb. In other words, we say that any system with
fewer than \sbb\ degrees of freedom is not counted as an observer; systems with
greater than \sbb\ degrees of freedom have a chance of being observers
if they also satisfy other properties which we will not examine here. \sbb\ is related to
the entropy of the object under consideration in that it is the logarithm of the number of states,
but we are interested in constructing ordered systems with \sbb\ degrees of freedom, so \sbb\ 
is not literally the entropy.

What is a reasonable estimate for \sbb? The number of degrees of freedom
 in a
person is about equal to the number of particles, so roughly we can
divide the mass of a person by the mass of the proton to get
\be
\sbb \sim 10^{30} ~.
\ee
Perhaps we only want to count entire civilizations living on planets
as observers. The earth has about $10^{22}$ more particles than a
person, so this estimate would give 
\be
\sbb \sim 10^{50}
\ee
Surely no more entropy than this is required to form intelligent
observers; the amount of intelligence per particle on the earth is
miniscule. On the other hand, it is quite possible that intelligent observers can be produced with far fewer particles than in a person, so we will summarize our ignorance by the $1\sigma$ estimate
\be
\sbb = 10^{35 \pm 15} \ \ \ \ \ \  (1 \sigma)~.
\ee

Now we can estimate the number of Boltzmann Brains formed in a given
 vacuum.
First of all, the particle physics of the vacuum may not allow
for the formation of interesting structures, in which case the number
of Boltzmann Brains  is zero. If particle physics allows for the formation of
interesting structures, the cosmological constant may be too large, so
that there is not enough room to make interesting structures. Finally,
if the cosmological constant is reasonably small and the particle
physics allows for interesting structures to form, we can estimate the
number of Boltzmann Brains which form.

In equilibrium, all of the entropy of \ds\ space is in the horizon. On
average, one graviton is present in the bulk. In
order to make a Boltzmann Brain, we must remove entropy from the
horizon and build an ordered structure. If this structure has a size
of order the Hubble scale, then the number of degrees of freedom in
the structure is about equal to the number of degrees of freedom
removed from the horizon. The Boltzmann Brain we are building is an
ordered state and therefore has a small entropy compared to the number
of degrees of freedom it contains. So, in order to build a \bb, we
remove \sbb\ degrees of freedom from the horizon and put them into an
ordered structure. This process decreases the entropy of the horizon by
\sbb; since the \bb\ has small entropy relative to the number of degrees of freedom,
the entire system decreases its entropy by about \sbb.
Therefore, the time to produce a \bb\ is given by 
\be
t_{BB} \approx H^{-1} e^{\sbb}~.
\ee
Note that this is actually a lower bound on the time to produce a \bb, because 
the particle physics of the vacuum may prevent ordered structures from forming efficiently.
For example, if the mass of the particles is large then extra energy must be expended in
building a \bb. Therefore the above argument really gives a rough bound,
\be
t_{BB} > H^{-1} e^{\sbb}~.
\ee

The expected number of \bb s produced in a given vacuum is
\be
N_{BB} = {t_{\rm decay} \over t_{BB}}
\ee
The decay time is given by the exponential of the instanton action
\be
t_{\rm decay}  \sim e^{S_{inst} }~.
\ee
It is helpful to make explicit the double-exponential nature of \tbb\ by defining
\be
t_{BB} \equiv H^{-1} e^{B_{BB}}~.
\ee
Our argument above gives
\be
\bbb  > \sbb ~.
\ee
Now the number of \bb s is given by
\be
N_{BB} = {t_{\rm decay} \over \tbb} \sim e^{S_{inst} - \bbb}
\ee
Recall that \bbb\ is an exponentially large number. 
Generically, $B_{BB}$
and $S_{inst}$ are not of the same order, so the exponent is dominated
by the larger of the two. Therefore, there are two regimes. If the
instanton action is smaller than \bbb , the number of \bb s
produced is double-exponentially small,
\be
N_{BB} \sim e^{- \bbb}~,
\ee
On the other hand, if the instanton action is larger than \bbb , 
so that the decay time is longer than the \bb\ time, then a
double-exponentially large number of \bb s are produced,
\be
N_{BB} \sim e^{S_{inst}} > e^{\bbb}
\ee
In any given vacuum, the number of Boltzmann Brains
produced is either essentially zero or double-exponentially
large.

\subsection{Summary} 
As we mentioned above, a method of regulating infinities is necessary before we can say that a 
double-exponentially large number of \bb s in one causal patch is a problem. We believe that a fair summary of the current situation is the following:
all proposed measures whose predictions are known and 
which are not already ruled out \cite{raphprob, scalefactor} require
that all vacua in the landscape decay before they produce \bb s,
\be
\label{bound}
\boxed{
t_{\rm decay} < t_{BB}~.}
\ee
See \cite{bf, lindebb, lindebbb, bfy, vilbb, bfytwo} for more detailed discussion.

For the sake of having a concrete number to think about, a wide class of vacua will be able 
to produce \bb s relatively efficiently. We estimated above
that in a vacuum with reasonably cooperative particle physics \tbb\ is simply related
to the number of degrees of freedom required for an intelligent observer,
\be
t_{BB} \approx H^{-1} e^{\sbb}~.
\ee
Basing our crude estimates for what constitutes an observer on ourselves, we found
\be
\sbb = 10^{35 \pm 15}
\ee
where the uncertainty represents our lack of knowledge of the appropriate definition of the minimal intelligent observer. Putting in some additional uncertainty to account for how efficiently different vacua can produce \bb s, a useful number to keep in mind, valid for a wide class of vacua, is
\be
\label{tbbest}
\boxed{
\tbb \approx H^{-1} e^{10^{40 \pm 20}}~.}
\ee
Although this time is absurdly large and absurdly uncertain, it is parametrically shorter than the recurrence time for vacua such as our own. Therefore the proposed bound is nontrivial.

\section{False Vacuum Decay}
\label{CDL_review}

Before focusing on our specific example, we point out that at the level of low-energy effective field theory coupled to gravity, it is easy to build false vacua which live for about the recurrence time.
The relevant formulae for metastable vacuum decay were described by Coleman and De Luccia (CDL) \cite{cdl}.  The CDL formalism computes the semi-classical tunneling rate from a Euclidean instanton 
that interpolates between the true and false vacua
in four-dimensional low-energy effective field theory coupled to gravity.

The CDL tunneling probability in the thin-wall limit, where the transition region between vacua in the instanton solution can essentially be treated as a domain wall, is a function only of the initial vacuum energy $V_i$, the final energy $V_f$, and the tension of the domain wall $\tau$.

The tunneling rate per unit four-volume is proportional to $e^{-B}$, where the bounce action $B = S_{\rm CDL}[\phi] - S_i$ is the difference between the actions of the CDL instanton and the background, is given by \cite{parke}
\be
\label{bounceParke}
B = \frac{27\pi^2\tau^4}{2(\delta V)^3} r(x,y)
\ee
where the first factor is the quantum field theory result and 
\be
\label{rdef}
r(x,y) = 2 \frac{1+xy - \sqrt{1+2xy+x^2}}{x^2(y^2-1) \sqrt{1+2xy+x^2}}
\ee
is the correction due to gravity, where\footnote{\cite{parke} and \cite{cdgkl} use different definitions of $x$ and $y$.  We follow the conventions of \cite{cdgkl}.}
\bea
\label{xdef}
x &=& \frac{\tau^2}{\tau_c^2} = \frac{3 G_4 \tau^2}{4\delta V} \\
\label{ydef}
y &=& \frac{V_i + V_f}{\delta V} \, .
\eea
The critical tension is defined as
\be
\tau_c = \sqrt{\frac{4\delta V}{3G_4}} \, ,
\ee
where $\delta V = V_i - V_f$ and $G_4$ is the four-dimensional Newton's constant.

The radius of the domain wall is given by extremizing the Euclidean action of the instanton; in the thin-wall limit, it is given by
\be
\label{radiusParke}
\rho = \frac{\rho_0}{\sqrt{1+2xy+x^2}}
\ee
where $\rho_0 = 3\tau/\delta V$ is the result from field theory and the denominator is a correction due to gravity.

We will focus on the case of a metastable dS with small vacuum energy $V_i =V_{dS} \gtrsim 0$ decaying to a true vacuum with negative energy $V_f = V_{AdS} < 0$.\footnote{When $V_f < 0$, the final state is not in fact eternal AdS but rather an open FRW spacetime, which collapses in a big crunch.}   The simplest estimate of the decay time, which is in fact an upper bound, is just the Poincare recurrence time.  As the  action of the CDL instanton $S_{CDL}[\phi]$ is negative, 
\be
\label{ratebound}
B \leq -S_i  = \frac{24\pi^2}{G_4^2 V_{dS}} \, .
\ee
This is basically the best estimate one can give, using only the one energy scale $V_{dS}$.  If, as in our universe, $V_{dS}$ is extremely small, the bound (\ref{ratebound}) on the decay rate is extremely weak.  

For improving this bound, we'll be interested in two particular limits of (\ref{bounceParke}). The behavior of the decay rate crosses over sharply as the tension $\tau$ crosses the critical tension $\tau_c$.  When the tension is subcritical, $x \ll 1$.  Expanding the square root in the numerator of (\ref{rdef}) to second order in $x$, the gravitational correction becomes
\be
r \approx  1 \,  ,
\ee
and the decay rate is given simply by the field-theory result
\be
B \approx \frac{27\pi^2\tau^4}{2(\delta V)^3} \, 
\ee
which is significantly smaller than the upper bound (\ref{ratebound}).  Gravity is negligible because the bubble size
\be
\rho \sim \sqrt{ \frac{x}{G_4 \delta V}} \sim \sqrt{ x} \, l_{AdS}
\ee
is much smaller than both the radius of curvature $l_{AdS}$ of the false vacuum and the dS radius $l_{dS}$, since $l_{AdS} \ll l_{dS}$ for the cases of interest to us.

The lifetime increases rapidly when the tension becomes critical, $x=1$, and the bubble radius reaches a maximum at $\rho = \sqrt{3/G_4V_{dS}} = l_{dS}$.  For supercritical tension, $x \gg 1$, the gravitational correction (\ref{rdef}) to leading order becomes
\be
r \approx \frac{2}{x^2 (y+1)} = \frac{16 (\delta V)^3}{9 G_4^2 \tau^4 V_i} \, .
\ee
Plugging this in to (\ref{bounceParke}) we find
\be
B \approx \frac{24\pi^2}{G_4^2 V_{dS}} 
\ee
which nearly saturates the bound (\ref{ratebound}), meaning the lifetime is approximately the Poincare time.  Again, the bubble radius $\rho \sim l_{AdS}/\sqrt{x}$ is small, but instanton spacetime is also very small and so the contribution to $B$ from $S_{CDL}$ is negligible.

\section{Decay rate of the KKLT construction}
\label{rate_simple}
In this section, we briefly review the geometry of the flux vacua used in the KKLT construction \cite{gkp, kklt}. We then compute the decay rate using the brane description of the instanton. We find that the lifetime cannot be made parametrically long and, in fact, is independent of the SUSY breaking
scale for long throats. 
We find that the lifetime is bounded by
$\exp(10^{22})$ Hubble times. The analysis in this section is
simplified in that we neglect certain corrections to the tension of
the domain wall mediating the decay. However, we present this
analysis first because the formula for the corrections is not
known with certainty. In the next section, we estimate the corrections and
find that they do not affect our conclusions.

\subsection{Geometry}
We start in type IIB string theory with D7 branes and O3 planes compactified on a $CY_3$, or
equivalently, an F-theory compactification on an elliptically fibered $CY_4$. 
Adding $F_3$ and $H_3$ fluxes generates a tree-level superpotential $W_0$; further
nonperturbative effects stabilize the volume modulus \cite{kklt}.

In the presence of fluxes, the compact manifold is a conformal
Calabi-Yau, so we can write the string-frame metric
\be
ds^2 = h^{-1/2}(y) g_{\mu \nu}(x)dx^\mu dx^\nu + h^{1/2}(y) e^{2u}
\hat g_{mn}(y) dy^m dy^n~.
\label{metric}
\ee
Here $\hat g_{mn}(y)$ is the fiducial Calabi-Yau metric on the
manifold, which we have defined so that
\be
\int d^6 y \sqrt {\hat g} = l_s^6~.
\label{hatmet}
\ee
Thus the unwarped volume of the compactification is
\begin{equation}
V_6 = e^{6u} l_s^6 
\end{equation}
We assume that near some point the Calabi-Yau looks like a deformed
conifold with deformation parameter $S$. The Calabi-Yau metric
$\hat{g}_{mn}$ in this region is approximately
\begin{equation}
d\hat s^2 \approx dr^2 + r^2 ds_{T_{1,1}}^2~.
\end{equation}
This metric is valid between a UV cutoff $r_0$ where the fact that the
CY is not simply a conifold becomes apparent and an IR cutoff
$\tilde{r} \sim S^{1/3}$ where the deformation becomes important. 
 The deformed conifold has two
holomorphic 3-cycles: the A-cycle, which is a 3-sphere with volume
$S$, and the B-cycle, which is noncompact in the conifold solution.  
Conifoldology \cite{gkp} relates the deformation
parameter to the fluxes through the cycles via\footnote{Many works define the parameter $z$ by
\be
z = \exp \left(- {2 \pi K \over g_s M} \right)
\ee
This parameter is related to our parameter $S$ by a factor of $r_0^3$;
this factor is often ignored.}
\begin{equation}
S = r_0^3 e^{-{2 \pi K \over g_s M}}
\end{equation}
where $M$ is the number of units of flux through the A-cycle and $K$
is the number of units of flux through the B-cycle:
\be
M = {1 \over (2 \pi l_s)^2 } \int_A F_3\ \ \ \  \ \ \ \ 
K =  {-1 \over  (2 \pi l_s)^2} \int_B H_3 ~.
\ee
$K$ depends on the UV cutoff $r_0$, but $S$ is a parameter of the
infrared physics and does not depend on the cutoff.

Between the UV cutoff $r_0$ and the IR cutoff $\tilde r$, the warp factor is approximately \cite{kt, ks}
\be
h = 1 + {L^4 \left[\log (r/\tilde{r}) + 1/4\right] \over r^4}
\ee
with
\be
L^4 = {81 (g_s M)^2 l_s^4 \over 8 e^{4u}} 
\ee
The exact metric in the throat is known (see, for example,
\cite{hko}); what will be important for us is that
the proper volume of the minimal A-cycle is
\be
V_{S^3} =2 \pi^2 (b g_s M)^{3/2} l_s^3
\ee
with the constant $b \approx 0.932 $.
$S$ gives the volume of the minimal $S^3$ in
the metric $\hat g_{m n}$, so we have a useful relation between the complex structure and the geometry, 
\be
h_{\rm tip}^{3/4} e^{3u} S = V_{S^3} 
\label{svrel}
\ee
where $h_{\rm tip}$ is the warp factor at the infrared end of the throat.

Note that the deformation parameter $S$ is exponentially small;
equation (\ref{svrel}) shows that that $h_{\rm tip}$ is exponentially large
and they scale as \cite{gkp}
\be
S \sim h_{\rm tip}^{-3/4}~.
\ee
It is these exponentially small parameters which allow us to
break supersymmetry by an exponentially small amount.

\subsection{SUSY breaking and decay rate}
\label{decayrate}
SUSY is broken by adding $N_\dtb$ anti-$D3$ branes at the tip of the
throat.
The  contibution of the \dtb s to the
action is
\be
S_\dtb = {N_\dtb \over (2 \pi)^3 g_s l_s^4} \int d^4x \, h_{\rm tip}^{-1}~.
\ee
We work in the string frame.
Due to the warp factor, from the 4d point of view, this looks like an exponentially small additional energy density,\footnote{The factor of 2 in $\delta V$ is explained in \cite{kpv}; half the energy comes from the tension of the \dtb 's and the other half from the potential energy in the $F_5$ field induced by the fluxes.}
\be
\label{deltaV}
\delta V = {2N_\dtb \over (2 \pi)^3 g_s l_s^4} h_{\rm tip}^{-1}
\ee

The $\b{D3}$s at the tip of the throat are subject to decay by the KPV
mechanism of brane/flux annihilation \cite{kpv, flw}.  The $\b{D3}$s sit near one pole
of the $S^3$ at the tip of the throat, but, due to the $H_5 = g_s^{-2}
\star_{10} H_3$ flux they polarize into an $NS5$ wrapping an $S^2$ of
the $S^3$.  If $M < 12N_\dtb$, the $NS5$ is large enough to slide around
the equator of the $S^3$ to the other pole where it de-polarizes into
$D3$s.  The $F_3$ and $H_3$ flux carry a $D3$-brane charge $MK$, which
in this process ``annihilates" with the $\b{D3}$s, leaving $K-1$
units of $H_3$ flux and $M-N_\dtb$ $D3$s, conserving 3-brane charge.  For
$M > 12N_\dtb$, the case relevant to metastable dS, the $NS5$ classically
sits near the original pole but can decay to the other pole via
tunneling across the equator.

In the thin-wall limit, which is a good approximation for small $N_\dtb $,
the instanton mediating the KPV decay is a Euclidean NS5 bubble at a
fixed radius in the 4d spacetime and wrapping the $S^3$.  The action of the 
NS5-brane wrapping the 3-sphere at the tip of the throat is 
\be 
S_{NS5} = {1 \over (2 \pi)^5 g_s^2 l_s^6} V_{S^3} \int d^3x \, h_{\rm tip}^{-3/4} 
\ee
From the 4d point of view the NS5-brane is just a domain wall
separating the interior true vacuum, with no \dtb s and SUSY
restored, from the exterior false vacuum where SUSY is broken and the \dtb s are present.
The tension of this domain wall is just the 4d effective tension of the NS5-brane 
\be
\tau_{NS5} = {1 \over (2 \pi)^5 g_s^2 l_s^6} V_{S^3} h_{\rm tip}^{-3/4}  
  = { b^{3/2}M^{3/2} \over 16 \pi^3 g_s^{1/2} l_s^3} h_{\rm tip}^{-3/4}~.
\label{nsten}
\ee

For now we will assume we can ignore gravitational corrections to the decay, and in the next section we'll check whether we can.  In the field theory approximation the instanton solution is given by just the tension  of the domain wall (\ref{nsten}) and the difference in vacuum energy (\ref{deltaV}).

The radius of the domain wall (\ref{radiusParke}) in the field theory approximation is
\begin{equation}
\label{bubble_radius}
\rho = {3 \tau \over \delta V} = {3 b^{3/2} M^{3/2}g_s^{1/2} \over 4 N_\dtb} l_s h_{\rm tip}^{1/4}
\end{equation}
and the action (\ref{bounceParke}) is\footnote{This matches the decay rate found by \cite{kpv} in 2001, up to the famous $(2 \pi)^7/4$ correction to the exponent made in version 4 from 2006 and an additional factor of $4$ correction to the exponent which we have discovered; our $b$ is their $b_0^2$, and our $N_\dtb $ is their $p$.}
\begin{equation}
\label{lt}
B_{\rm KPV} = S_{\rm CDL} =  {27 \pi^2 \over 2} {\tau^4 \over (\delta V)^3} = {27 b^6 
  \over 2048 \pi} {g_s M^6 \over (N_\dtb) ^3} ~.
\end{equation}
The warp factor \htip\ has cancelled out! Although the warped geometry
allows for an exponentially small SUSY breaking scale, the decay rate
is actually independent of the amount of warping. Note that also the
volume of the compactification has cancelled out, so that the lifetime
depends only on $g_s$ and the amount of flux $M$.

There is an intuitive explanation for why the warp factor cancels out
of the decay rate. The entire decay process is localized near the tip
of the throat; the \dtb s which provide the difference in vacuum
energy are localized at the tip, and $NS5$ brane which mediates the
decay is also localized at the tip. So the entire process is
insensitive to how far away the bulk of the Calabi-Yau is. In fact,
the only reason the warp factor appeared at all is that we are
measuring quantities relative to the bulk. For processes localized at
the tip, we can write everything in terms of proper quantities which
are then independent of the warp factor:
\bea
\tau_{\rm proper} = h^{3/4} \tau &=& { b^{3/2}M^{3/2} \over 16 \pi^3 g_s^{1/2} l_s^3}  \\
\delta V_{\rm proper} = h \delta V &=& {2 N_\dtb \over (2 \pi)^3 g_s l_s^4} \\
\rho_{\rm proper} = h^{-1/4} \rho &=&  {3 b^{3/2} M^{3/2}g_s^{1/2} \over 4 N_\dtb} l_s
\eea
This makes it clear that the instanton action cannot depend on the warp factor, at least
in the field theory approximation.\footnote{Although we are not computing the one-loop determinant here, a similar argument would tell us that the decay time will depend on the warp factor in such a way that the proper
decay time is independent of the warping, so we get
\be
t_{\rm decay} \sim h_{\rm tip}^{1/4} \exp \left({27 b^6 
  \over 2048 \pi} {g_s M^6 \over (N_\dtb) ^3} \right)~ .
\ee
It is not completely clear that this argument is correct, but in any case the exponential
gives the dominant behavior in the regime of interest.}

Now, having computed the decay rate, we can deduce a maximum lifetime.
Plugging in the value of $b$ in equation (\ref{lt}), we find
\be
\boxed{B_{\rm KPV} \approx 3 \cdot 10^{-3} {g_s M^6 \over (N_\dtb)^3}}
\ee
How big can this quantity possibly be? First, we set $N_\dtb=1$ to make $B$ 
as large as possible. The tadpole constraint coming from the conservation of $F_5$ flux is
\be
M K < {\chi \over 24}
\ee
where $\chi$ is the Euler number of the $CY_4$ of the F-theory compactification.  In addition, consistency of the warped compactification requires the deformation parameter $S$ to be exponentially small, which implies $K > g_sM$.  Combining this with the tadpole constraint, we obtain
\be
g_s M^2 < {\chi \over 24}
\label{tad}
\ee
Furthermore, requiring that the minimal $S_3$ be bigger than the
string scale gives the additional constraint
\be
\label{largetHooft}
g_s M > 1~,
\ee
which when combined with (\ref{tad}) gives a maximum for $M$,
\be
\label{Mbound}
M < {\chi \over 24} ~.
\ee
Thus the instanton action is bounded by
\begin{equation}
\label{BKPVbound}
B_{\rm KPV} <  3 \cdot 10^{-3} g_s M^2 M^4 < 
 4 \cdot 10^{-10} \chi^5 
\end{equation}
yielding a bound on the decay time 
\be
t_{\rm decay} < \exp \left( 4 \cdot 10^{-10} \chi^5 \right)~.
\ee
Since we have not computed the one-loop determinant we do not know the dimensional prefactor
which should appear in front of the exponential. Since the decay is a field theory process the prefactor
is likely to be a microphysics length scale which is much shorter than the Hubble length. This allows
us to write
\be
\boxed{t_{\rm decay} < H^{-1} \exp \left( 4 \cdot 10^{-10}  \chi^5 \right)}~.
\ee
It is not known whether $\chi$ has a finite upper bound; the existence of a bound has neither been proven nor disproven.  Examples of elliptically fibered $CY_4$'s with Euler number $\chi$ up to around $10^{6}$ have been found \cite{lsw}.   Assuming a bound near this value exists, the lifetime is can be bounded roughly by
\begin{equation}
t_{\rm decay} < H^{-1} \exp \left(10^{22} \right)~.
\end{equation}
Even if no geometrical bound on $\chi$ exists, there may be physics considerations which limit its size.

Note that since our bound on the lifetime depends exponentially on $\chi^5$, it is extremely sensitive to the largest possible $\chi$. It is fascinating that the largest known $\chi$ agrees so well with the bound given by (\ref{bound}) and (\ref{tbbest})  coming from \bb\ considerations.

\subsection{Gravitational corrections}
\label{gravitational_corrections}

We have just used the field theory limit to compute the decay
rate. Now we must check whether gravitational corrections are really
unimportant. This is more than a technicality because, as we reviewed in section \ref{CDL_review}, it is gravitational corrections which can make the lifetime of order the recurrence time. 

The \ds\ vacua we are considering must have very nearly
zero cosmological constant to have a chance of having a lifetime of order
$\exp(10^{40})$ because the lifetime is always bounded by the recurrence time. 
We achieve a small de Sitter cosmological constant by tuning $W_0$ to almost cancel the
uplifting term from the \dtb s. The supersymmetric AdS minimum,
however, will not have an extraordinarily small cosmological constant, because 
we want the supersymmetry breaking scale to be larger than the scale set by the \ds\ 
cosmological constant, and the supersymmetry breaking scale is related to the amount
of uplifting $\delta V$.

As discussed above in section \ref{CDL_review}, gravitational corrections are negligible when $x \ll 1$ which, for $\delta V\approx V_{AdS}$, is essentially equivalent to $\rho \ll l_{AdS}$.  Plugging (\ref{nsten}), (\ref{deltaV}), and
\begin{equation}
G_4 = {G_{10} \over V_6} = {(2\pi)^7 g_s^2 l_s^8 \over 2 e^{6u} l_s^6}
\end{equation}
in (\ref{xdef}), we find
\be
x = \frac{3 \pi^4 b^3 g_s^2 M^3}{2 e^{6u} N_\dtb} h_{\rm tip}^{-1/2} \, .
\ee
Note that here the warp factor does not cancel; warping is very
effective in limiting the gravitational backreaction because the
energy of the process is small compared to bulk scales.

Although the presence of the warp factor $h_{\rm tip}^{-1/2}$ means that 
that gravitational corrections can easily be made very small, we want to know if
the gravitational corrections are big for any reasonable choice of parameters.
Demanding that the total volume of the compactification is
bigger than the volume in the throat gives a bound \cite{giantinf}
\be
\label{volmin}
e^{4u} > 3 \pi^3 g_s M K \, .
\ee
Recalling that we need $K> g_s M$, this becomes
\be
e^{4u} > 3 \pi^3 g_s^2 M^2 \, .
\ee
In addition, $N_\dtb \ge  1$, so the gravitational corrections are bounded by 
\be
x< {b^3\over 2 \sqrt{3\pi} g_s} h_{\rm tip}^{-1/2} \approx 0.1 g_s^{-1} h_{\rm tip}^{-1/2}
\ee
Combining the inequalities (\ref{largetHooft}) and (\ref{Mbound}), we find the lower bound on the string coupling to be
\be
g_s > \frac{24}{\chi}  ~.
\ee
Assuming, as before, that $\chi$ is bounded by its maximum known value of around $10^6$, we now have
\be
x< 10^4 h_{\rm tip}^{-1/2} \,.
\ee
This quantity can be bigger than one for acceptable, although not extremely natural, choices of parameters, so we need to worry about gravitational corrections.
Holding fixed the parameters $g_s$ and $M$ which control the field theory decay rate, dialing the warp factor controls the strength of the gravitational corrections.  At the microscopic level, this corresponds to dialing the flux $K$ through the B-cycle, which controls the length of the throat as well as the deformation parameter $S$.  If $h_{\rm tip} > 10^8$, the gravitational corrections are indeed small, and we can safely use the field theory result (\ref{lt}) and rely on the bound (\ref{BKPVbound}).  For fixed $g_s$ and $M$, therefore, brane/flux annihilation in long, large-$K$ throats occur at field-theory rates.

On the other hand, if $h_{\rm tip}$ is too small, gravitational corrections are large.  As discussed in section \ref{CDL_review}, when $x >1$ the tension is supercritical and the decay rate nearly saturates the recurrence bound (\ref{ratebound}):
\be
\label{BKPVsupercritical}
B_{\rm KPV} \approx \frac{24 \pi^2}{G_4^2 V_{dS}}~~!
\ee
For short, small-$K$ throats, brane/flux annihilation therefore occurs extremely slowly.  However, in the regime where the gravitational corrections are important, supersymmetry is also badly broken. We will see in the next section that other decay modes will become important in this regime. 

\subsection{Destabilization of Bulk Fluxes}
\label{sectioncdgkl}
Having computed the decay rate via brane/flux annihilation, we consider whether it is really the dominant decay mode.  Without a completely detailed description of the vacuum, it is impossible to be sure the fastest decay has truly been identified.  However, all flux vacua have a very generic decay mode whose rate can be estimated.

 Recall that in the bulk we
have wrapped fluxes on a variety of cycles.  Before supersymmetry is
broken, there are BPS domain walls, branes wrapped on cycles
which can interpolate between vacua with different flux configurations.  
And, of course, with unbroken supersymmetry there are no instabilities.

However, if supersymmetry is broken by a small amount by uplifting to a dS vacuum, some of these now near-BPS domain walls become the bubble walls of instantons mediating genuine instabilities.  Ceresole, Dall'Agata, Giryavets, Kallosh, and Linde \cite{cdgkl} estimated the decay rate in precisely these circumstances.  To first order in the size of the SUSY breaking, the bubble size and decay rate depend only on the change in vacuum energies and not on the change in tension.  Therefore, the bubble tension can be approximated by the tension of the associated BPS domain wall.  For a supersymmetric AdS, with vacuum energy $V_{AdS}$, uplifted to slightly positive cosmological constant, $V_{dS} \ll |V_{AdS}|$, the bounce action is approximately
\be
\label{rateCDGKL}
B_{\rm CDGKL} = \frac{6\pi^2}{G_4^2 |V_{AdS}|} 
\ee
where $|V_{AdS}|$ is also approximately the size of SUSY breaking. 

We will first consider the case when $x < 1$ and gravitational corrections are unimportant.  To compare the rate (\ref{rateCDGKL}) to the decay rate by brane/flux annihilation (\ref{lt}), it is helpful to multiply and divide by the radius $\rho_0$ of the critical bubble for the KPV decay in the field theory approximation,
\be
\rho_0 \sim {\tau \over \delta V}~,
\ee
to get 
\be
B_{\rm CDGKL} \sim  {\ell_{AdS}^4 \over \rho_0^4 }(V_{AdS} \rho_0^4) \sim {\ell_{AdS}^4
  \over \rho_0^4 } B^0_{\rm KPV} \, .
\ee
where $B^0_{\rm KPV}$ is the action for the brane/flux annihilation in the field theory approximation. 
Recall that the quantity $ {\rho_0^4 \over \ell_{AdS}^4
  } \sim x^2$ controls the gravitational corrections. So we can write
\be
B_{\rm CDGKL} \sim  \frac{B^0_{\rm KPV}}{x^2}
\ee
For $x < 1$, $B_{CDGKL} > B^0_{\rm KPV}$, so the destabilization of bulk fluxes is slower than the brane/flux annihilation, and since gravity is unimportant the instanton action is well approximated by the field theory result $B^0_{\rm KPV}$. 

On the other hand, when $x > 1$ and gravity is important, the brane/flux annihilation rate instead approaches the recurrence rate (\ref{BKPVsupercritical}).  However, in this regime
$B_{\rm CDGKL} < B^0_{\rm KPV}$, so the destabilization of bulk fluxes is the most important process and the decay is even faster than the field theory approximation to the KPV decay.

Thus we can summarize the instanton action by
\bea
B &=& B^0_{\rm KPV} \ \ \ \ \ \ \ \ \ \  x<1 \\
B &\sim& \frac{B^0_{\rm KPV}}{x^2} \ \ \ \ \ \ \ \ \ x > 1 
\eea
Therefore up to possible order one factors in the exponent, the decay rate is bounded by
\be
t_{\rm decay} < \exp\left(B^0_{\rm KPV}\right)
\ee
so our simple analysis from the previous section gives the correct bound.

While the estimate of \cite{cdgkl} is the best estimate for the decay rate of nearly
supersymmetric vacua of which we are aware, there may well be constructions which are
longer lived than this estimate. In particular, \cite{cdgkl} assumes that before supersymmetry breaking some of the BPS domain walls have exactly the critical tension, so that the decay is just marginally forbidden. This assumption is not always correct for BPS domain walls, as mentioned by \cite{cdgkl}. A construction which is more stable under supersymmetry breaking than the estimate of \cite{cdgkl} could well provide a counterexample to 
our proposed bound.

\subsection{Summary}
To summarize, we have bounded the decay rate of the metastable KKLT vacuum.  In the regime of long throats, the dominant decay is by brane/flux annihilation and warping has no effect on the decay rate.  For short throats, the decay is instead by decay of bulk fluxes whose rate is given by (\ref{rateCDGKL}).  The lifetime, which depends simply on the flux $M$ wrapped on the $S^3$ at the tip of the throat and the string coupling $g_s$, is
\be
t_{\rm decay} \sim e^{3 \cdot 10^{-3} g_s M^6 /N_\dtb^3}~.
\ee
A computation of the one-loop determinant would be necessary to determine the dimensional factor
multiplying the exponential.

Putting in the tadpole constraint, demanding that the supergravity
approximation is at least marginally valid, and arguing that the dimensional prefactor is small
compared to the Hubble scale $H^{-1}$, we get a bound 
\begin{equation}
t_{\rm decay} < H^{-1}e^{4 \cdot 10^{-10} \chi^5} < H^{-1}e^{10^{22}}
\end{equation}
where to get the second inequality we have assumed that the Euler number is bounded by 
$\chi < 2 \cdot 10^6$.

Our result appears to depend sensitively on details, and we urge other authors to try to violate the bound in different constructions. Our bound depends sensitively on the largest possible $\chi$, which is not known. Additionally, it relies heavily on the formula (\ref{rateCDGKL}) to estimate certain decays, and the formula may not be generally true.  Finally, our estimates are valid for supersymmetry breaking by anti-$D3$ branes; the lifetime could be much longer for other types of supersymmetry breaking. Nevertheless, even within our simplified context the fact that the lifetime satisfies the bound proposed in \cite{bf} is nontrivial and surprising.

\section{Corrections to the tension}
\label{corrections}
In the previous section, we approximated the tension of the domain
wall by the tension of the wrapped $NS5$ brane. In fact, there are
other contributions to the tension of the domain wall; these
contributions could have the effect of increasing the lifetime. For
example, the parameter $S$ which controls the deformation of the conifold changes
in the transition; taking this into account increases the bounce
action. In fact, the additional action due to the change in $S$
appears to be the dominant correction. 

This correction was first computed by Frey, Lippert, and
Williams \cite{flw}. Here we will review that computation, updating
it to reflect an improved understanding of the Kahler
potential and correcting some minor errors which arose due to conflicting conventions in the literature.  
However, these results remain uncertain because computing
the correct Kahler potential in warped compactifications remains an
open problem; see \cite{gm, dst, dt, stud}. 

The contribution to the action from the closed string moduli is
naturally computed from the 4D superpotential in
the 4D Einstein frame.  Therefore, in contrast to the section \ref{rate_simple} 
where we worked in string frame, in this section we work in the 4D Einstein frame.

To compute the full tension, including the effect of the closed string
moduli, we use the an approximation similar to that of \cite{cdgkl} as described in section \ref{sectioncdgkl}.  
We have been interested in describing the brane/flux annihilation which leads to the decay of the
\dtb s. We can compute the tension by relating the domain wall we are
interested in to a BPS domain wall. Even in the absence of \dtb s,
one can consider a wrapped $NS 5$ brane domain wall. On one side of
the domain wall we have fluxes $K$ and $M$, and on the other side we
have fluxes $K-1$ and $M$ along with $M$ explicit $D3$ branes. This
is essentially the same domain wall which changes the flux through the
B-cycle by one unit, but now both sides are supersymmetric and we can
compute the tension using the BPS formula
\be
\tau_E = \left|\Delta \left( e^{\kah/2} W \right) \right|
\ee
where the notation $\tau_E$ indicates that this is the tension
computed in the 4D Einstein frame. It is unclear to us whether our calculation is exact for BPS domain walls or not, due to the complications associated with Kahler moduli in warped compactifications.

This supersymmetric domain wall does not constitute an
instability. If we now add a small number of \dtb s, we expect that
the tension computed from the BPS formula will not change much, but
now the domain wall interpolates between a nonsupersymmetric false
vacuum and a supersymmetric true vacuum, and the corresponding
instanton describes a real instability. In the following we compute the
tension in the supersymmetric case.

The superpotential is
\be
W = W_{\rm flux} + W_{\rm np}
\ee
We choose the following set of conventions
\bea
W_{\rm flux} = {1 \over (2 \pi)^7 l_s^8} \int G \wedge \Omega  \\
\int \Omega \wedge \bar{\Omega} = l_s^6 \\
e^{\kah/2} = {g_s^2 l_s^6 \over V_w}
\eea
where $G = F - \tau H$ with $\tau = i/g_s$, $V_w$ is the warped volume of the compactification, and $\Omega$ is the holomorphic three-form.

The flux superpotential can be evaluated by using the formula
\be
\int G \wedge \Omega = \sum_i \left( \int_A^i G \int_B^i \Omega  -  \int_B^i G \int_A^i \Omega \right)
\ee
where the sum is over all symplectic pairs of three-cycles.
For the conifold throat we have
\bea
\int_A G = (2 \pi l_s)^2 M \ \ \ \ \ \ \ \ \ \ \int_B G &=& {i (2 \pi l_s)^2 \over g_s} K \\
\int_A \Omega = S \ \ \ \ \ \ \ \ \ \ \ \ \ \ \ \ \ \ \  \int_B \Omega &=& {1 \over 2 \pi i} S \left(\log {S \over r_0^3 } - 1 \right)
\eea
Plugging these in, we get a formula for the contribution of the throat to the superpotential,
\be
W_{\rm throat} = -{i \over (2 \pi)^5 l_s^6} \left[ {K \over g_s} S +
 {M \over 2 \pi } S \left(\log{S \over
  r_0^3} - 1\right) \right]
\ee
Evaluating this at the supersymmetric minimum $D_S W \approx
\partial_S W_{\rm throat} =0$, we get\footnote{In \cite{flw} it is claimed that ${W_{\rm throat } \vline}_{vac}  = 0$ because the
$K$ and $M$ fluxes are $(2,1)$ forms. Our explicit calculation here
gives a nonzero answer, which does not depend on UV physics, and is equal to what
one would get from the field theory analysis, so we believe this
answer is correct. The conflict is resolved as
follows. In the noncompact conifold the fluxes $are \ (2,1)$ forms,
but because the manifold is noncompact this is not sufficient to
conclude that  ${W_{\rm throat } \vline}_{vac}  = 0$. Once the conifold is
embedded in a compact Calabi-Yau, it is no longer clear that the
fluxes are $(2,1)$ forms.}
\be
{W_{\rm throat } \vline}_{vac} = {i \over (2 \pi)^6 l_s^6} M S~.
\ee
Assuming that the change in the superpotential and Kahler potential is
small and that $g_s$ does not change much in the transition, the tension is
\be
\tau_E \approx  g_s^2 l_s^6 \left| { \Delta W \over  V_w} - {\Delta V_w \over  V_w^2} W
\right|
\ee
As pointed out by, for example, \cite{dst, stud}, 
although in these conventions the unwarped volume is
independent of the complex structure moduli, the warped volume is not.

Across the domain wall, $K$ decreases by one unit while M stays
fixed. The change in superpotential is therefore
\be
\Delta W ={i \over (2 \pi)^6 l_s^6} M \Delta S 
\ee
Recall that $S = r_0^3 \exp [-2 \pi K/(g_s M)]$, so
\be
\Delta S = {2 \pi \over g_s M} S
\ee
assuming that $2 \pi /(g_s M) \ll 1$, as it should be in the
supergravity approximation.
Then
\be
\Delta W = {i \over (2 \pi)^5 l_s^6 g_s} S ~.
\ee

Computing the change in the warped volume across the
domain wall is subtle, because on the side with $K-1$ units of flux
through the B-cycle there are $M$ explicit $D3$ branes. If one ignores
the backreaction of the $D3$ branes on the metric, then one finds that
the change in the warped volume has a strange UV dependence. One can
do the calculation correctly by finding the full metric with the $D3$
branes included, but the answer can instead be estimated by the following intuitive argument. 
Across the domain wall, one step in the Klebanov-Strassler
cascade has been eliminated. The change in warped volume is just the
warped volume of the eliminated region. So, we just need the warped
volume of the last step of the Klebanov-Strassler cascade. 

Since this argument will not get order one factors right, we will not
keep them here. The proper AdS radius in the IR is $\ell_{IR} = e^u L \sim (g_s M)^{1/2}
l_s$. We can compute the warped volume of this step:
\be
\Delta V_w \sim \int_{\tilde{r}^K}^{\tilde{r}^{K-1}} \ell_{IR}^6 {dr
  \over r} h_{\rm tip}^{-1/2}
\ee
The first part is the proper volume, and to get the warped volume we
multiply by the factor $h_{\rm tip}^{-1/2}$. The relationship between
the IR cutoffs $\tilde r^K$ and $\tilde r^{K-1}$ is
\be
{\tilde r^{K-1} \over \tilde r^K}= \left({S^{K-1} \over S^K}\right)^{1/3}
= e^ {2 \pi \over 3g_s M}
\ee
Performing the integral, we
get
\be
\Delta V_w\sim \ell_{IR}^6  h_{\rm tip}^{-1/2} {1 \over g_s M} \sim (g_s M)^2
h_{\rm tip}^{-1/2}  l_s^6 \, .
\ee
One can perform this analysis in the full warped deformed conifold metric and get the same result.

Gathering together the above formulas we get
\be
\tau_E = g_s^2 l_s^6 \left| {1 \over (2 \pi)^5 l_s^6}{ S \over
    g_s V_w} + 
c {(g_s M)^2 l_s^6\over V_w^2}
 h_{\rm tip}^{-1/2} W
\right|
\ee
where $c$ is an unknown order one constant into which we have absorbed the relative phase between the two terms.

We would like to compare this formula to the tension we computed from
the probe $NS 5$ brane computation. To translate, we must relate $S$
to the geometrical factors appearing in the $NS 5$ computation.  From (\ref{svrel}), the
parameter $S$ is the size of the $S^3$ at the tip of the conifold with
the Kahler modulus and the warp factor factored out. To get the
physical volume we put these back in:
\be
 S = h_{\rm tip}^{-3/4} e^{-3u} V_{S^3} .
\ee
Using the approximation that the warped volume of the compactification is about the same as the unwarped volume, $e^{6u}l_s^6 \approx V_w$, and rearranging some factors, we get
\be
\tau_E = {g_s^3 l_s^9 \over  V_w^{3/2}} \left|  {1 \over
   (2 \pi)^5 g_s^2 l_s^6} V_{S^3} h_{\rm tip}^{-3/4} + c {g_s M^2 l_s^3 \over
    V_w^{1/2}} W h_{\rm tip}^{-1/2} \right|
\ee
This is the tension computed in the Einstein frame. The prefactor is
precisely the conversion from string frame to Einstein frame, so we
drop this in comparing to our formula from the probe $NS 5$
computation.  The first term inside the absolute value is precisely the  wrapped $NS 5$ brane
tension, equation (\ref{nsten}).  The second term can be thought of as the contribution to the action
due to changing the closed string moduli. It is suppressed by
additional powers of the volume and factors of $g_s$. However, the
warp factor at the tip $h_{\rm tip}$ is exponentially large, and the
second term is suppressed by fewer powers of $h_{\rm tip}$. Therefore it, and not the
tension of the wrapped $NS 5$ brane, could be the dominant
contribution for a wide range of parameters.

For us, however, this term will not be important. The reason is that
we are interested in a situation where the nonsupersymmetric vacuum
has nearly zero cosmological constant. This requires $V_{AdS} + \delta V \approx 0$ which implies 
\be
W \sim 
h_{\rm tip}^{-1/2} \, .
\ee
Thus for uplifting to nearly flat space the correction term in the tension becomes
\be
\Delta \tau \sim 
h_{\rm tip}^{-1}
\ee
which is now smaller, in terms of powers of $h_{\rm tip}$, than the first
term; with some more work one can see that in fact the correction is
always negligible. Therefore, we are justified in using the tension
calculated from the probe $NS 5 $ brane calculation.

We have assumed in the above that the string coupling $g_s$ and the
volume modulus $\sigma$ do not change significantly across the domain
wall. One can compute the additional contribution to the action from
these terms and find that it is not important in the regime of interest.

\section{Delicacy of the KKLT construction}
\label{KKLTproblems}

Upon investigating the parameter space of controllable KKLT dS vacua, 
we discover that in fact stabilizing the
volume with nonperturbative corrections to the superpotential and then
breaking supersymmetry with \dtb s is not easy to control. The basic
tension is that large flux numbers in the throat are desirable so that
supergravity is valid and the nonsupersymmetric vacuum is
metastable. On the other hand, large flux numbers in the throat make
the volume of the compactification large. However, the nonperturbative
corrections to the superpotential are exponentially small at large
volume. It is challenging to find parameters for which the volume is
large enough to allow metastable nonsupersymmetric vacua but small
enough so that the nonperturbative volume stabilization mechanism can
work.

The compact volume has to be large enough so that the throat fits.  In terms of the 
imaginary part of the universal Kahler modulus, 
equation (\ref{volmin}) can be restated as \cite{giantinf}
\be
\sigma > 3 \pi^3 M K 
\ee
where the $\sigma =  g_s^{-1}V_w^{2/3}$.

More generally, we need some room for other cycles wrapped with fluxes
so that we can tune $W_0$, so the requirement is actually
\be
\sigma = 3 \pi^3 M K \left({V_6 \over V_{\rm throat}}\right)^{2/3}~.
\ee
where $V_6$ is the volume of the compact manifold and $V_{\rm throat}$ is
the volume of the throat region.
Warping is not significant in this formula because both the warped volume and
the unwarped volume of the throat are dominated by the region near the bulk
where the warp factor approaches one. 

The warped solution requires $K > g_s M$, and
in order that the \dtb s are perturbatively stable against brane/flux annihilation, we need \cite{kpv}
\be
M > 12 N_\dtb \, .
\ee
Also, the radius of the minimal $S_3$ is given by $\sqrt{ b g_s M}$,
so for the supergravity solution to be reliable we need $g_s M \gg 1$.

To make use of these inequalities, we rewrite the formula for the volume modulus as
\be
\sigma = 36 \pi^3 N_\dtb \left( M \over 12 N_\dtb \right)  (g_s M) \left( K \over g_s M \right)  
\left({V_6 \over V_{\rm throat}}\right)^{2/3}
\ee
The volume modulus is roughly $10^3$ times a number of factors, each of which must
be larger than one by the arguments above. One would have been tempted to make each one of
these factors large in order to obtain control.

Such a large volume may be difficult to obtain in the KKLT
construction because nonperturbative effects must be important.
More quantitatively, the superpotential is
\be
W = W_0 + A e^{- a \sigma}
\ee
and the Kahler potential is
\be
\kah = - 3 \log \sigma + ...
\ee
so solving $D_\sigma W =0$ for the supersymmetric vacuum we get
\be
W_0 = -{a A \sigma \over 3} e^{-a \sigma} 
\ee
We want to know how large $\sigma$ can be subject to solving this
equation. The smallest $|W_0|/\sigma^{3/2}$ is about $1/\sqrt{N_{vac}}$, or perhaps
$10^{-2000}$ \cite{denef}. This gives roughly
\be
a \sigma - \log A < 5000
\ee
If the nonperturbative effects come from gaugino condensation on $D7$
branes, then $a = 2 \pi/N_{D7}$. As far as we know, an extremely large
number of $D7$ branes is not possible, so we assume that $a >
0.1$. Then, if the prefactor $A$ does not take an extreme value, we
have
\be
\sigma < 10^5~.
\ee
which leaves an extremely narrow window where the construction can
work,
\be
 10^3 \, N_\dtb \, \left( M
  \over 12 N_\dtb \right)  (g_s M) \left( K \over g_s M \right)
 \left({V_6 \over V_{\rm throat}}\right)^{2/3}
 <
 \sigma < 10^5
\label{window}
\ee
Recall that each of the factors on the left side of the equation must be larger
than one. 
In the words of S. Kachru, constructions in this narrow window 
``are not deep in the regime of calculability.''\cite{kach}

There may be ways to arrange for $\sigma$ to take a larger volume than
our estimate of $10^5$.
As pointed out by Denef et al. \cite{ddfgk}, the prefactor
 $A$ may be quite large,
\be
A \sim e^{2 \pi \chi(D) \over 24 g_s}
\ee
where $\chi(D)$ is the Euler number of the divisor $D$ on which the $D7$s are wrapped.
Also, one can impose a discrete $R$-symmetry so that $W_0$ is zero at
tree level \cite{kachrev, choi}; this would allow for a much smaller minimum value of
$W_0$. 
This latter possibility has recently been explored in more detail \cite{bmp}, and has the
advantage that all of the analysis in this paper remains valid in computing the decay rates.

Of course, the large volume scenario of \cite{largevol} allows for much
larger volumes, but in this case supersymmetry is already broken when
the moduli are stabilized, so we would have to do an entirely
different estimate of the decay rates.

Finally, one could perhaps avoid the need for such large volumes by
breaking supersymmetry in a milder way than by adding antibranes. Note
that it is only the combination of volume stabilization by
nonperturbative effects and supersymmetry breaking by antibranes which
squeezes us into the narrow window (\ref{window}).

\section{Conclusions and Future Directions}
\label{conclusions}
We have investigated a new bound stating that all de Sitter vacua should decay before they produce Boltzmann Brains. This time scale is much longer than the Hubble time but much shorter than the recurrence time for vacua with small cosmological constant such as our own. We have found surprisingly strong support for the bound in a sector of the landscape, the KKLT vacua, in which one might have thought it would be easy to construct very long-lived vacua.
Incidentally, we have pointed out that the classic KKLT construction is quite difficult to control. However, we expect that minor modifications can lead to much more controlled de Sitter vacua.
Our analysis has narrowly focused on the specific example of KKLT vacua, but
we suspect that this type of bound may be an example of a phenomenon generic to stringy dS vacua.
It would be of great interest to see whether other constructions of de Sitter space obey the same bound, since our results appear to be highly model-dependent.

The basic reason that all de Sitter vacua might decay before they make Boltzmann Brains is that stabilizing moduli and tuning the vacuum energy to be small requires a rich set of ingredients. Since the vacuum energy is accidentally small, the ingredients in the construction will naturally have 
decay rates which are unrelated to the scale of the vacuum energy. Also, we have seen that nearly supersymmetric vacua are not necessarily extremely stable. In the case of the KKLT vacua, we have found the decay rate is actually independent of the supersymmetry breaking scale.

On the other hand, it is quite possible that by considering a slightly different construction, other authors will be able to construct extremely long-lived vacua. In this case, the currently viable measures
would be ruled out, and
we would have valuable new information about the correct way to regulate the infinities of eternal inflation. Finally, it would be very interesting to find a model-independent argument which bounds the lifetimes of de Sitter vacua without invoking Boltzmann Brains.

\section*{Acknowledgements}
We particularly thank Raphael Bousso for collaboration in the early
stages of this project and Shamit Kachru for discussions and 
technical assistance. We have also enjoyed helpful discussions with Chris Beem,
Steve Giddings, Maximilian Kreuzer, Andrei Linde, Liam McAllister, Yu Nakayama, Stephen Shenker, Eva Silverstein, 
and Leonard Susskind. 
This work was supported by Israel Science Foundation grant 568/05,
 the Berkeley Center for Theoretical Physics, and by DOE grant DE-AC0376SF00098.

\end{document}